\documentclass[epj]{svjour}
\usepackage{amssymb}  
\usepackage{graphicx}

\providecommand{\tr}{\mathrm{tr}\,}
\renewcommand{\phi}{\varphi}
\renewcommand{\d}{\mathrm{d}}

\begin{document}
%
\title{Foams in contact with solid boundaries: equilibrium conditions and conformal invariance}
\author{M. Mancini and C. Oguey
} 
%
%
\institute{LPTM\thanks{CNRS UMR 8089}, Universit\'e de Cergy-Pontoise, 95031 Cergy-Pontoise, France\\ \email{mancini@ptm.u-cergy.fr \textnormal{and} oguey@ptm.u-cergy.fr}}
%
\date{v3.5 \today / Received: date/ Revised version: date}
%
\abstract{
A liquid foam in contact with a solid surface forms a two-dimensional foam on the surface.
We derive the equilibrium equations for this 2D foam when the solid surface is curved and smooth, generalising the standard case of flat Hele Shaw cells. The equilibrium conditions at the vertices in 2D, at the edges in 3D, are invariant by conformal transformations. Regarding the films, conformal invariance only holds with restrictions, which we explicit for 3D and flat 2D foams. Considering foams confined in thin interstices between two non parallel plates, normal incidence and Laplace's law lead to an approximate equation relating the plate profile to the conformal map. Solutions are given for the logarithm and power laws in the case of constant pressure. The paper concludes on a comparison with available experimental data.
%
\PACS{
      {82.70.Rr}{Aerosols and foams}  \and
      {82.70.Kj}{Emulsions and suspensions}  \and
      {68.03.Hj}{Gas-liquid and vacuum-liquid interfaces: Structure, measurements and simulations}
     } 
} 
\maketitle
%

\section{Introduction} \label{secIntro}

Many applications of liquid foams, from dish washing to shaving cream, container decontamination, penetration and dispersion through porous media like rocks, etc., involve contacts with solid surfaces. 

One of the most common settings, for both experimental and theoretical investigations, is the Hele Shaw cell consisting of two flat and parallel plates separated by a small distance. The foam, confined in the interstice, may often be considered as quasi bi-dimensional; the equilibrium equations are 2D versions of Plateau's equations  \cite{WeaireRiv,WeaireHut}.
Even in this simple setting, the predictions of straightforward dimensional reduction may fail when the thickness of the bubbles are not much smaller than their diameter in the plane of the plates. Then instabilities occur \cite{WeaireInstab,Brakke,Fomenko}.

In the present paper, we consider standard dry liquid foams in contact with smooth curved solid surfaces. Later, we will come back to settings similar to the Hele Shaw cell. For the solid surface ---the container wall in typical applications---, the main assumptions are rigidity and smoothness. Contact with non rigid walls such as elastic membranes or an interface with another fluid medium must be treated differently; in a number of cases, at least for statics, flexible surfaces may be included as part of the foam with a surface tension differing from the rest of the foam. Smoothness rules out corners, wedges as well as dirty or fractal surfaces. Usually, the films are pinned by such irregularities.

On a smooth and clean surface $S$, the contacts of the real 3D foam form a 2D foam in $S$. Upon meeting the boundary surface, the thin films, or cell interfaces, define edges in $S$, the Plateau borders (edges in 3D) end up onto vertices in $S$ and the cells are incident to faces, the 2D cells in $S$. The films are free to move along the surface, with viscous dissipation in dynamic conditions and full relaxation to equilibrium in static conditions. 

In recent experiments, the simple parallelepiped geometry was
modified. Whereas the focus for hemispherical, yet parallel, plates
was on topology \cite{DiMeglioSenden}, Drenckhan et al. \cite{Drenck} tried several non parallel chambers and noticed that each resulting 2D foam could be related to the reference regular honeycomb pattern by a conformal map. 
The dominant constraint was volume conservation. A question addressed in the present paper is whether the shape of the films, that is, the local geometry of the foam, can also be related to the conformal map and how. 

Conformality means that the transformation locally reduces to a simple isotropic dilatation, preserving angles. 
However, in 2D, general conformal transformations are not compatible with Laplace law
(at least, the 2D version).
The only compatible conformal transformations (mapping circle arcs to circle arcs) are the Moebius transformations (homographies) \cite{Moukarzel,Weaire,ManciniO}. In the case of algebraic functions (with exponents other than $\pm 1$), or the logarithm, or any other holomorphic function, is conformality only a long wavelength approximation ? 
or is there a correction to the 2D equilibrium ?

On one hand, the equilibrium conditions at the Plateau borders ---of dimension 2 less than the embedding space, the edges in 3D, the vertices in 2D--- turn out to be invariant by conformal mappings. This result is exact within the standard models of dry foams in any dimension.
On the other hand, regarding the films ---bubbles interfaces, of co-dimension 1--- in thin cells, we will show that incorporating the curvature in the third direction 
restores compatibility with Laplace, implying a relation between the profile and the map.
The case of constant pressure will be solved explicitly.
Comparison with constant volume settings, including experimental ones, will also be done.

\emph{Outline}.
After recalling some basics on foam equilibrium and contact with a solid boundary in Section \ref{secEquil}, we derive, in Section \ref{secEquilBoundary}, the equilibrium equations at the boundary in the case of normal incidence. Section \ref{secConform} establishes conformal invariance for the co-dimension 2 Plateau borders equilibrium. Section \ref{secConformFilm} treats the question of conformal mappings for the films and solves specific examples to lowest order in non parallelism. 
Section \ref{secConclusion} concludes.
Appendix \ref{apCurves} and \ref{apCurvatures} recall elements of geometry for non experts.
\ref{apNonNormalInc} contains some extensions dealing with the cases where the surface tension along the solid varies from cell to cell, so that the contact angle is not right. 

\section{Foam equilibrium} \label{secEquil}
In equilibrated dry foams, the bubbles are separated by films obeying Laplace-Young's law:
\begin{equation}\label{ulaplace}
  \Delta P + 2\gamma H =0
\end{equation}
where $\Delta P=P_2-P_1$ is the pressure drop, $\gamma$ the surface tension and $H$ the mean curvature of the film.
In static conditions and in absence of applied field, the pressure inside each bubble is constant so that the films are constant mean curvature surfaces.

The films and cells meet three by three at edges in a manner satisfying Plateau's laws \cite{WeaireHut} :
\begin{eqnarray}
  \label{uplateaub}&\sum_{j=1}^3\gamma_j \mathbf{b}_j=0\\
  \label{uplateauk}&\sum_{j=1}^3\gamma_j H_j=0
\end{eqnarray}

The first law asserts that the sum of the forces on any edge is zero; indeed the force per unit edge length exerted by the film $\mathcal{F}_j$ is the surface tension $\gamma_j$ times the unit vector $\mathbf{b}_j$ tangent to the film and normal to the junction.
In general situations, the surface tension may have different values $\gamma_{j}$ on different sides.

The second law (\ref{uplateauk}) is the sum of the pressure drops along a small circular path encircling the edge. By (\ref{uplateaub}), this sum of differences is zero as long as the path is closed: $(P_1-P_2)+(P_2-P_3)+(P_3-P_1)=0$. Notice the sign convention implicitly assumed under (\ref{uplateauk}): given a sense of circulation, the pressure difference in going from bubble 1 to bubble 2 is defined as $P_2-P_1$ (head minus tail). By (\ref{ulaplace}), this locally fixes the orientation of the three films around any given edge or Plateau border. The same film $\mathcal{F}$ may appear with different signs in sums around different edges.

\subsection{Contact with a boundary}\label{secBoundary}

The foam is in contact with a solid wall, not necessarily flat but smooth and clean. 
The films or interfaces can freely slide
along the surface so as to relax to equilibrium. 

With these assumptions, an interface $i$ has a contact with the wall only in non wetting conditions; then the contact angle $\theta$ satisfies Young's law:
\begin{equation}\label{young}
  \gamma_{j} \cos \theta + \gamma_{S1} - \gamma_{S2} =0
\end{equation}
where 1 and 2 label the two fluids (gas in normal foams) on both sides of the film.

In the case of a film ---a double locally symmetric interface--- the two bubbles contain the same gas so that $\gamma_{1S} = \gamma_{2S}$ and $\cos \theta = 0$. This condition of \emph{normal incidence} is typical of soap froth; it is also valid when the soap solution wets the solid surface, provided the wetting films on both sides have similar surface tension.

In the standard Hele-Shaw experimental setup, when the glass plates separation $h$ is small, smaller than any typical length (curvature radius or cell diameter) of the 2D foam, the curvature $k_\perp$ of the films in the direction normal to the plates is approximately zero.  
Then, the 2D foam satisfies equilibrium equations similar to (\ref{uplateaub}), (\ref{uplateauk}) with quantities adapted to 2D: line tension $\gamma_j$, edge tangent vector $\mathbf b_j$ and curvature $H_j=(2r_j)^{-1}$, $r_j$ being the signed curvature radius of film $i$. In flat equilibrated 2D foams, the edges are circle arcs.

Our purpose, next, is to consider non planar glass surfaces.

\section{Equilibrium conditions at a boundary}\label{secEquilBoundary}

We derive equilibrium equations, analogous to Plateau's laws, valid for the 2D foam formed by the contacts of the films with the solid boundary.

\begin{figure} \centering
  \includegraphics[width=0.99\columnwidth]{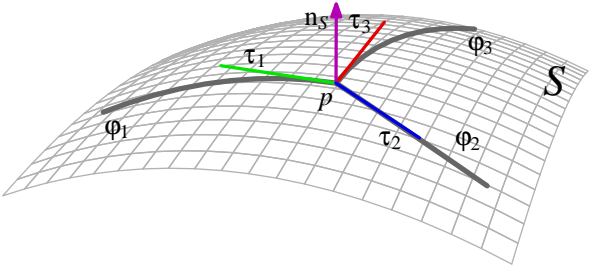}
  \caption{The neighbourhood of a vertex $p$ of a 2D foam: three films and a Plateau border
    meet at $p$ on the solid surface $\mathcal{S}$.}
  \label{Svert-neighborhood}
\end{figure}

A Plateau border is a line $\psi$ at the junction of three  films $\mathcal{F}_j, j=1,2,3$.
We call this a \emph{triple}.
These films and border are in contact with a solid smooth surface $\mathcal{S}$.
The contact of each film $\mathcal{F}_j$ gives rise to a line $\phi_j$ in $S$; these three lines emanate from a vertex $p$ of the 2D foam that is the contact of the border $\psi$ with $\mathcal S$ (fig. \ref{Svert-neighborhood}). Unless otherwise specified, all the curves are parametrised by arc length $s$ with $s=0$ at $p$. Then, for any curve $\phi$, $\tau=\dot{\mathbf\phi}=\frac{d}{ds}\phi$ is the unit vector field tangent to $\phi$.

Assume normal incidence on $\mathcal{S}$ as implied by equilibrium for real soap froth on clean surfaces. Then
\emph{
At $p$, common to $\mathcal{F}_1,\,\mathcal{F}_2,\,\mathcal{F}_3$ and $\mathcal{S}$, 
the following equations are verified:
\begin{eqnarray}
  \label{2Dplatot}&\sum_{j=1}^3\gamma_j \tau_j=0,\\
  \label{2Dplatok}&\sum_{j=1}^3\gamma_j k_g(\phi_j,\mathcal{S})=0,
\end{eqnarray}
where $k_g(\phi_j,\mathcal{S})$ is the geodesic curvature of $\phi_j$ in $\mathcal{S}$.
}

The geodesic curvature is defined in several books (e.g. \cite{Morgan,Willmore}) and in Appendix A.

When all the film tensions are equal, $\gamma_j = \gamma$, the triangles representing the vector sums in (\ref{uplateaub}) and (\ref{2Dplatot}) are equilateral and the Plateau angles are $2\pi/3$. 

For the proofs, we use two lemmas.

\subsection{Two lemmas}

The following lemmas are valid in full generality. The normal incidence assumption will only be used to prove the main equations (\ref{2Dplatot}), (\ref{2Dplatok}).

\begin{lemma} \label{lemCurvatures}
Let $\mathcal{F},\mathcal{S}$ be two smooth embedded surfaces intersecting transversely\footnote{The surfaces are nowhere tangent to each other.} on a line $\phi$.
Then the geodesic and normal curvatures satisfy the following relation at any intersection point $x$:
\begin{equation} \label{eqlemma}
  \left(\begin{array}{c} k_g(\phi,\mathcal{S})\\
      k_n(\phi,\mathcal{S})\end{array}\right)
  =\left(\begin{array}{cc}
       \cos\theta & -\sin\theta\\
       \sin\theta & \cos\theta
    \end{array}\right) 
  \left(\begin{array}{c} k_g(\phi,\mathcal{F})\\
      k_n(\phi,\mathcal{F})\end{array}\right)
\end{equation}
where $\theta$ is the angle between the two surfaces at $x$.
\end{lemma}

On an edge $\psi$, the tangent vector $\dot\mathbf\psi$, the unit normal $\mathbf n_j$ (short for $\mathbf n_{\mathcal F_j}$) and the co-normal $\mathbf b_j$ form an orthonormal basis for each $i$. The circular sign convention implies the following:
\begin{lemma} \label{lemPlatoEdge}
On a Plateau border $\psi$ the first equilibrium equation (\ref{uplateaub}) is equivalent to 
\begin{equation}\label{platoPerp}
  \sum_{j=1}^3 \gamma_j \mathbf{n}_j =0.
\end{equation}
In turn, (\ref{platoPerp}) implies
\begin{eqnarray}\label{psiDdot}
\label{platoKn}&\sum_{j=1}^3\gamma_j k_n(\mathbf\psi, \mathcal{F}_j)=0\,,\\
\label{platoKg}&\sum_{j=1}^3\gamma_j k_g(\mathbf\psi, \mathcal{F}_j)=0\,.
\end{eqnarray}
Moreover the pair (\ref{platoKn}), (\ref{platoKg}) is equivalent to (\ref{platoPerp}) at all points where $\psi$ is curved.
\end{lemma}

\subsection{Proofs}

\subsubsection{Proof of  Lemma \ref{lemCurvatures}}

In the plane perpendicular to $\phi$ at $x$, the two orthonormal bases 
$(\mathbf{n}_F, \mathbf{n}_F \wedge \dot\mathbf\phi)$ and 
$(\mathbf{n}_S, \mathbf{n}_S \wedge \dot\mathbf\phi)$ are rotated by $\theta$ with respect to each other. Equation  (\ref{eqlemma}) is the coordinate change formula applied to the curvature vector $\mathbf{k}$ (Appendix \ref{apCurvatures}).
\qed

\subsubsection{Proof of Lemma \ref{lemPlatoEdge}}
The vector product of (\ref{uplateaub}) with $\dot\mathbf\psi$ gives (\ref{platoPerp}).

One gets (\ref{platoKn}) and (\ref{platoKg}) by taking the scalar product of (\ref{platoPerp}) with, respectively, the curvature vector $\mathbf k=\ddot\mathbf\psi$ and $\dot\mathbf\psi\wedge\mathbf k$ and applying the definitions of normal and geodesic curvatures (\ref{eqcurvcurv}). When $\mathbf k \neq 0$, the reciprocal is also guaranteed because $(\mathbf k, \dot\mathbf\psi\wedge\mathbf k)$ constitutes an orthogonal basis of the plane perpendicular to $\dot\psi$ where the vectors in (\ref{platoPerp}) are always confined.
\qed

\subsubsection{Proof of equation (\ref{2Dplatot})}

Because of normal incidence, the tangent $\dot\phi_j(0)$ coincides with the co-normal $\mathbf b_j:=\mathbf b(\psi, \mathcal{F}_j)$ at $p$.
Thus (\ref{2Dplatot}) is just a rewriting of (\ref{uplateaub}).
\qed

\subsubsection{Proof of equation (\ref{2Dplatok})}

By normal incidence again, $\dot\mathbf\psi$ and $\dot\mathbf\phi_j$ form an orthonormal basis in the tangent plane $T_p\mathcal{F}_j$. Therefore, as trace of the curvature operator (\ref{eqTrace}), 
$$ H_j=\frac{1}{2}\left(k_n(\phi_j, \mathcal{F}_j)+k_n(\psi, \mathcal{F}_j)\right).
$$
The linear combination with coefficients $\gamma_j$ yields
$$ \sum_{j=1}^3 2\gamma_jH_j=\sum_j \gamma_jk_n(\phi_j, \mathcal{F}_j)+\sum_j \gamma_j k_n(\psi, \mathcal{F}_j).
$$
The left hand side is zero by Laplace (\ref{uplateauk}). The last term on the right vanishes by (\ref{psiDdot}). Normal incidence, (\ref{eqlemma}) with $|\theta|=\pi/2$, gives 
$k_g(\phi_j,\mathcal{S}) = -k_n(\phi_j,\mathcal{F}_j)$ and concludes.
\qed

\section{Conformal invariance}\label{secConform}

Conformal invariance is a remarkable property of the models of equilibrated foams.
Of course, the exact symmetry group depends on the dimension and on the model, that is, on the set of equations fulfilled at equilibrium.
The equations may be divided into two sets: \textit{i}) the film, or cell boundaries, equation (\ref{ulaplace}) (Laplace-Young); \textit{ii}) Plateau's laws (\ref{uplateaub}, \ref{uplateauk}), or their 2D counterparts  (\ref{2Dplatot}, \ref{2Dplatok}), for the Plateau borders, which are the co-dimension 2 skeleton of the cellular complex.

In standard foams, without any externally applied field, the films are constant mean curvature surfaces (CMCS). In general, this property is not preserved by conformal mappings of the embedding space or region.

When external fields, such as gravity or electromagnetic forces, act on the foam, the films may not be CMCS anymore. This is also the case when the embedding space is not flat, as for a foam on a curved wall. Indeed, the pressure difference is the sum of two curvatures: the geodesic (longitudinal) one and the normal one, which may both vary along the curve even if the sum $H$ is constant. This is the more general situation considered here. If only the second set of equations is kept (Plateau), then the symmetry group is the group of conformal transformations. Let us treat the higher dimensional case first.

\subsection{Conformal invariance of the Plateau borders equilibrium.} \label{sec3DConform}
In Euclidean spaces of dimension larger than 2, any conformal transformation is the composition of similarities (translations, rotations, dilatation) and inversions (Liouville Thorin theorem \cite{Dubrovin}).

In a dry 3 or $D$-dimensional foam, consider a triple made of a Plateau border $\psi$ and three films $\mathcal{F}_j$, $j=1,2,3$. Then
\emph{%
the line equilibrium equations (\ref{uplateaub}) and (\ref{uplateauk}) are preserved by conformal transformations of space.
}

Our proof is similar to that of \cite{Weaire}.
By definition of conformality, equation (\ref{uplateaub}), stating that the dihedral angles between the films $\mathcal{F}_j$ at $\psi$ are constant, is invariant by conformal transformations.

Next, let us treat (\ref{uplateauk}). In Liouville Thorin's decomposition, the similarities obviously preserve (\ref{uplateaub}) and (\ref{uplateauk}), so we only need to check invariance for inversions.
Let $\vec{r}_j\,\in\,\mathbb{R}^3$ be the position of a point in $\mathcal{F}_j$ and $H_j$ the mean curvature at that point. Under the inversion $\vec{\tilde x}=\frac{\vec{x}}{x^2}$, the mean curvature transforms to $\tilde H_j$ given by \cite{Willmore}:   
\begin{equation}
\label{trasfH}
\tilde H_j= -H_j r_j^2-2\, \vec{r}_j \cdot\vec{n}_j.
\end{equation}
Then for any point $\vec{r}$ of $\tilde \psi$, the image of $\psi$, thus common to all the $\tilde\mathcal{F}_j$, the linear combination with coefficients $\gamma_j$ yields
\begin{equation}
\label{invaHsum}
\sum_{j=1}^3\gamma_j \tilde H_j=-r^2\sum_{j=1}^3\gamma_j H_j -2\,\vec{r}\cdot\sum_{j=1}^3\gamma_j \vec{n}_j.
\end{equation} 
The terms on the right hand side of (\ref{invaHsum}) vanish by (\ref{uplateauk}) and (\ref{platoPerp}) respectively, proving the claim.
\qed

Obviously, the tetrahedral figures around equilibrated vertices in 3D space are conserved by angle preserving transformations.

In 3 and higher $D$-dimensional foams, the conformal group of similarities and inversions is a symmetry of the entire foam, including the films satisfying Laplace's equation, only if the films are spherical caps; this is a very restricted subclass.

In the 2D Euclidean plane, however, any CMCS is a line of constant curvature, that is, a circle. Then, indeed, the symmetry group of 2D foams is  SL$(2,\mathbb{C})/\{1,-1\}$, the group of linear fractional transformations (or homographies or Moebius transformations) \cite{Moukarzel,Weaire,ManciniO}. The two sets of equations, including Laplace, are preserved in this case. Foams in standard flat Hele Shaw chambers fulfil all these conditions.

\subsection{Conformal invariance of bi-dimensional foams}\label{sec2DConform}

Specific to 2D is the existence of a much larger set of conformal transformations. This larger set is a symmetry of foams as characterised by the Plateau set of equations only.
The geodesic curvatures $k_j$ are not necessarily constant, the graph edges are of any smooth shape, but nodal equilibrium (Plateau conditions) is still assumed.

A further extension is that the degree (or coordination) $d$ of the vertices is arbitrary (but finite), whereas standard 2D foams have degree $d=3$.

The arguments leading to conformal invariance are detailed in
\cite{ManciniEuf04}. For consistency, the main steps are reported here. The embedding plane may be identified to the complex plane: $\mathbb{R}^2 \simeq \mathbb{C}$. In the plane, a foam is a planar graph.

\subsubsection{Equilibrium in 2D}
The 2D equilibrium conditions (\ref{2Dplatot}, \ref{2Dplatok})  at the vertices may be rewritten as
\begin{eqnarray}
\label{eqPlateau2Dt} \sum \gamma_{j}\, \vec{\tau}_{j}&=&0,\\
\label{eqPlateau2Dk} \sum \gamma_{j}\, k_{j}&=&0,
\end{eqnarray} 
where $k_j$ are the film geodesic curvatures and $\vec{\tau}_{j}$ the unit vectors tangent to the films. The sums are over the $d$ edges incident to the considered vertex.

\subsubsection{Conformal maps} \label{Conformal_maps} \label{secPathcomplex}

Two dimensional conformal maps are given by complex holomorphic
functions $f$ of the variable $z=x^1+ix^2$ in $\mathbb{C} \simeq
\mathbb{R}^2$. Indeed, Cauchy-Riemann's equation,
\begin{equation} \label{eqCauchyR}
\partial_{z^*} f = \frac{1}{2}(\partial_1+i\partial_2)(f^1+if^2) = 0,
\end{equation}
is equivalent to the conformality condition on the Jacobian matrix.
The complex derivative is written $f' = \partial_{z} f$.

The scalar product in the plane is $x\cdot y = \mathrm{Re}(x^* y) = \mathrm{Re}(x\, y^*)$ for any complex $x,y$.

\subsubsection{Paths under conformal maps} \label{Paths_transformations}

The edges of the graphs are described by parametrised $C^2$ curves
$ \phi: [t_1,t_2] \rightarrow \mathbb{R}^2$ (or $\mathbb{C}$),\quad $t
\mapsto \phi(t)$. The definitions of the local basis $(\tau, n)$ and
curvature $k$ are recalled in appendix \ref{apCurves}.

By a conformal $f$, a path $\phi$ is mapped to an image $\tilde{\phi}: 
[t_1,t_2] \ni t \mapsto \phi(t) \stackrel{f}{\mapsto} \tilde{\phi}(t) = f(\phi(t)) \in \mathbb{C}$.
The transformation rules are
\begin{eqnarray} \label{eqPathTransf2}
\dot{\tilde{\phi}} &=& f'\, \dot{\phi}\\
\ddot{\tilde{\phi}} &=& f'\, \ddot{\phi} + f'' (\dot{\phi})^2\\
\tilde{\tau} &=& \frac{f'}{|f'|}\, \tau\ ; \quad    \label{eqCTransformTau}
\tilde{n} = \frac{f'}{|f'|}\, n\\
\tilde{k} &=& \frac{\tilde{n}\cdot\ddot{\tilde{\phi}}}{|\dot{\tilde{\phi}}|^2}
 = \frac{1}{|f'|} \left(k + \mathrm{Re}\left[-i \frac{f''}{f'} \tau \right] \right),
   \label{eqCTransformK}
\end{eqnarray}
where $f'\equiv f'(\phi(t))$, etc.

\subsubsection{Conformal invariance of nodal equilibrium} \label{Conformal_invariance}

Let $X=(V,E,\gamma)$ be a planar graph with vertex set $V$, edges $E$ (given as twice differentiable curves) and line tensions $\gamma : E \rightarrow \mathbb{R}$. 

\emph{
Let $f$ be a conformal map of the plane.
Then the equilibrium conditions (\ref{eqPlateau2Dt}, \ref{eqPlateau2Dk}) are satisfied at all vertices of $\tilde{X} = f(X)$ if and only if they are satisfied by $X$.
}

Indeed, any vertex $x$ of the graph is a point common to all its incident edges $j=1,\ldots,d$. 

The equivalence for (\ref{eqPlateau2Dt}) is an immediate consequence of conformality as,
by (\ref{eqCTransformTau}), $\sum \gamma_{j}\, \tilde{\tau}_{j}=  \frac{f'(x)}{|f'(x)|} \sum \gamma_{j}\, \tau_{j}$.

For (\ref{eqPlateau2Dk}), we may use (\ref{eqCTransformK}):
\begin{equation} \label{eqEquivL}
\sum_{j=1}^d \gamma_{j}\, \tilde{k}_{j}=
 \frac{1}{|f'|} \left(\sum \gamma_{j}\, k_{j} + 
  \mathrm{Re}\left[-i \frac{f''}{f'} \sum \gamma_{j}\, \tau_j \right] \right),
\end{equation}
where everything is evaluated at $x$. By (\ref{eqPlateau2Dt}), the real part on the right vanishes so that the two sums, around $\tilde{x}$ and $x$, are proportional.
\qed

Drenckhan et al. \cite{Drenck} gave an alternative argument, based on the known invariance of Plateau's laws under homographies and Taylor expansion to second order.

\subsubsection{Generalisation} \label{Generalisation}

The embedding plane $\mathbb{C}$ may be replaced by analytic (Riemann) surfaces.
Invariance and the first proof still hold true in these more general cases where $X$ is a foam in a first surface $M$, $\tilde{X}=f(X)$ is the image of $X$ by a conformal map to a second surface: $f: M \rightarrow \tilde{M}$.

\section{Mapping the edges: foams between non parallel plates} \label{secConformFilm}

In the previous section, we have seen that the equilibrium conditions at the Plateau borders were invariant under conformal maps. Here, on a specific experimental set-up, we consider the films more closely. 

Among all the planar conformal maps, only the homographies, or Moebius transforms, map circles to circles. In the traditional setting of the Hele-Shaw cell,
normal incidence to the plates implies that the film curvature in the vertical direction (normal to the plates) is zero. Whence the mean curvature $H$ reduces to half the geodesic curvature of the trace of the films on the plates, which is a line. By Laplace, constant pressure in each of the bubbles implies that the geodesic curvature is constant along any bubble edge, therefore the edge is a circular arc.

Under mappings such as the logarithm or any other one not equal to a homography, the image of a circle arc is not a circle arc. So the resemblance of the pictures with the complex map may be questioned.

When the foam is sandwiched between plates that are not parallel, the films must bend in the direction normal to the plates. This slightly modifies the pressure balance as we now examine.

\subsection{Thin foam between two non parallel plates} \label{secNonpara}

\begin{figure}  \centering
  \includegraphics[height=2.0cm]{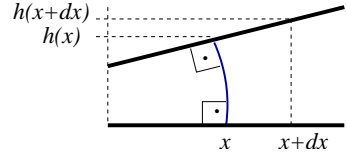}
  \caption{A film joining two non parallel plates.}
  \label{tilt}
\end{figure}

The upper plate ---more precisely its face wet by the foam--- is described by a height function $h(x^1,x^2)$ defined in the plane at the bottom of the cell. Both $h$ and its slope $|\nabla h|$ are supposed to be small. As a two dimensional vector, $\nabla h$ locally indicates the direction of maximal slope. The slope in this direction is the local dihedral angle (which is small). Because the film meets both top and bottom faces orthogonally, it must be curved: to lowest order, its sectional, or normal, curvature in the vertical direction is $k_v\simeq -n_\mathcal F \cdot \nabla h / h$ (fig. \ref{tilt}).
Within the same degree of approximation, we get the mean curvature $H$ by adding the horizontal sectional curvature, which nearly is the geodesic curvature $k:=k_g(\phi, \mathcal S)$ of the curve $\phi=\mathcal F\cap \mathcal S$ in the upper or lower boundary $\mathcal S$. Then, Laplace's law says
\begin{equation} \label{eqLapcor}
P_1-P_2 = 2 \gamma H_\mathcal{F}
= \gamma \left(k + k_v \right)
\simeq \gamma \left(k - \frac{n_\mathcal F \cdot \nabla h}{h} \right).
\end{equation}
Thus, in equilibrium, with pressure difference constant along the film, the geodesic curvature differs from a constant by the last term in (\ref{eqLapcor}), representing contributions from curvatures in the third direction.

Next, let us explicit the deformation from circularity.
Suppose the foam $X$ is conformally related to a reference one by $\tilde X=f(X)$.
Any edge $\phi$ has an image $\tilde \phi=f\circ \phi$ in the reference $\tilde X$.
Using $n=i\tau$, let us solve formula (\ref{eqCTransformK}) of for the curvature vector $k$:
\begin{equation}   \label{eqkn}
k= |f'| \tilde k + \mathrm{Re}\left[n\, (\ln f')' \right].
\end{equation}

In our approximations, $n_\mathcal F\simeq n$. Fusing (\ref{eqLapcor}), translated into complex notations, and (\ref{eqkn}) gives
\begin{equation} \label{eqfh}
(P_1-P_2)/ \gamma \simeq 
  |f'| \tilde k + \mathrm{Re}\left[n\, \partial_z(\ln f' - 2\ln h) \right].
\end{equation}
This equation, valid on any edge of the foam, is a constraint relating the conformal map $f$ and the cell profile $h$; it involves the edge through the pressure drop, the curvature of the reference edge and the unit normal.

For an equilibrated foam containing gas, the pressure drop on the LHS should be constant along any edge. Now, except for special cases, making the RHS constant seems to imply an intricate relation between the geometry of the edges, the mapping $f$ and the profile $h$.

Interesting cases are when the mapping $f$ leaves some freedom to the choice of the foam. For example, global motion by translation and rotation, eventually with appropriate deformation. These criteria are met when $P_1-P_2$ and $\tilde k$ are both identically zero. Let us treat some of these cases.

\subsection{Constant pressure foams} \label{secConstP}

For specificity, we assume that the reference foam is in a standard parallel Hele Shaw cell. Then the curvature $\tilde k$, on the right, is proportional to the pressure difference in the reference: $\tilde P_1-\tilde P_2 = \tilde\gamma \tilde k$.

When, moreover, the pressure is constant in both the curved and
reference foams, equation (\ref{eqfh}) reduces to 
$$\mathrm{Re}\left[n\, \partial_z(\ln f' - 2\ln h) \right] \simeq 0.
$$
If the bubbles are small compared to the length scale over which $f$ and $h$ vary, then the 2D vector $\partial_z(\ln f' - 2\ln h)$ must vanish. As $\ln f'$ is analytic and $\ln h$ real, a little reasoning leads to the condition $\ln h/h_0 = \mathrm{Re}\left[\ln f' \right]$, or
\begin{equation} \label{eqSolMin}
h/h_0 \simeq |f'|
\end{equation}
where $h_0$ is positive a constant.

\subsubsection*{Examples}
\begin{figure}  \centering
  \includegraphics[height=5cm]{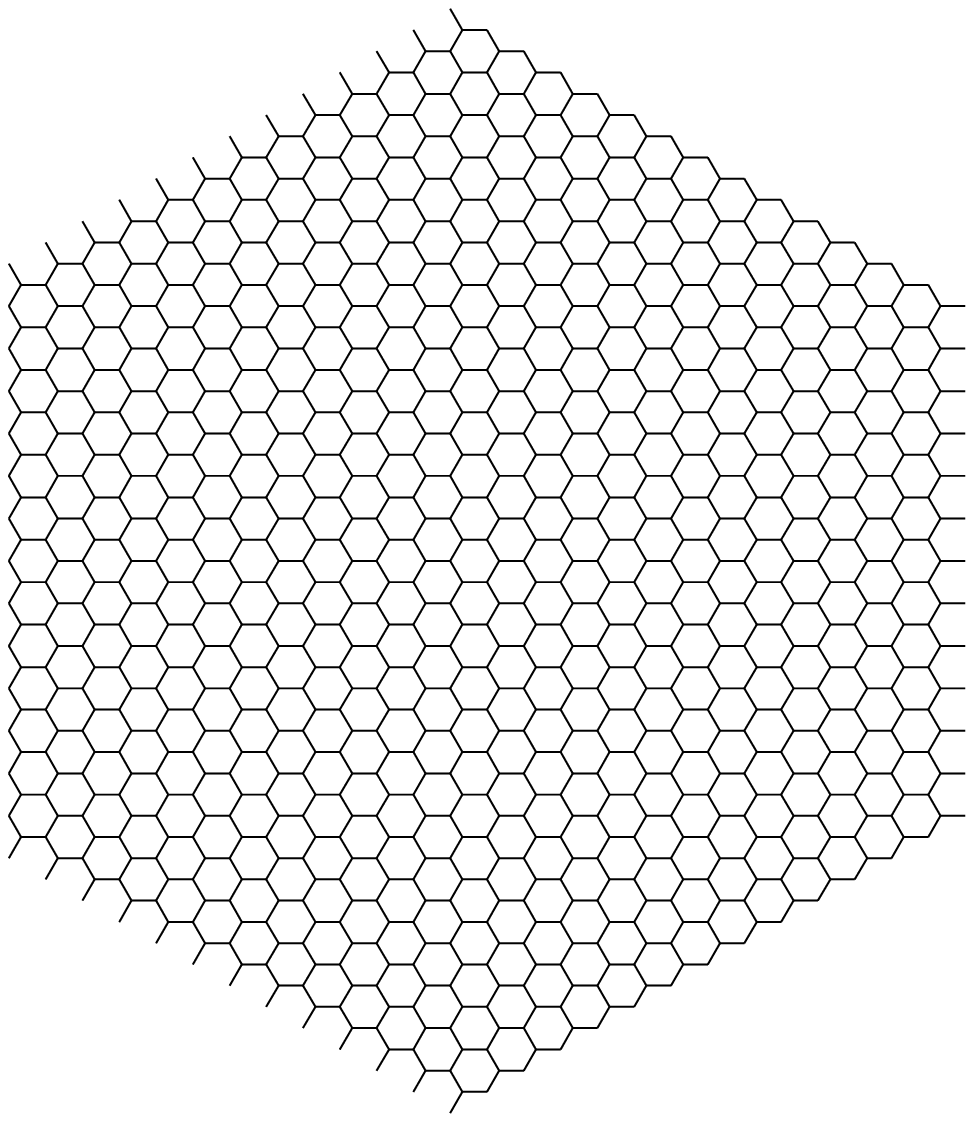}
  \caption{The reference regular hexagonal foam.}
  \label{figHexag}
\end{figure}

Following Drenckhan et al. \cite{Drenck}, we will most often take the hexagonal foam as reference (fig. \ref{figHexag}).

\subsubsection{Log map} \label{secLogedge}

The logarithmic map 
\begin{equation} \label{eqLogdef}
F(\tilde z) = \frac{1}{a^*} \ln(a^* \tilde z),
\end{equation}
where $a$ is a complex number or 2D vector, maps the outside of the unit disc to the upper half plane when $a$ is purely imaginary. This is one of the examples experimentally demonstrated in \cite{Drenck}. This mapping is also involved in gravity arches observed in ferro-fluid suspensions under gravity \cite{RothenPierRivJo,RothenPier} and in ferro-fluid bi-dimensional foams \cite{Elias}. It was observed that 'ln' is the only conformal map translationally invariant \cite{RothenPierRivJo,RothenPier,RivierConf}.

Applying (\ref{eqSolMin}) to the inverse mapping 
$f(z)=F^{-1}(z) = \frac{1}{a^*} \exp(a^* z)$,
the solution for the profile is
\begin{equation} \label{eqLogh}
h(z)\propto |f'| = \exp(\mathrm{Re}[a^*z]) = \exp(a\cdot z).
\end{equation}
This exponential profile
matches the experimental one, at least within the observable precision.

\begin{figure}  \centering
  \includegraphics[height=4cm]{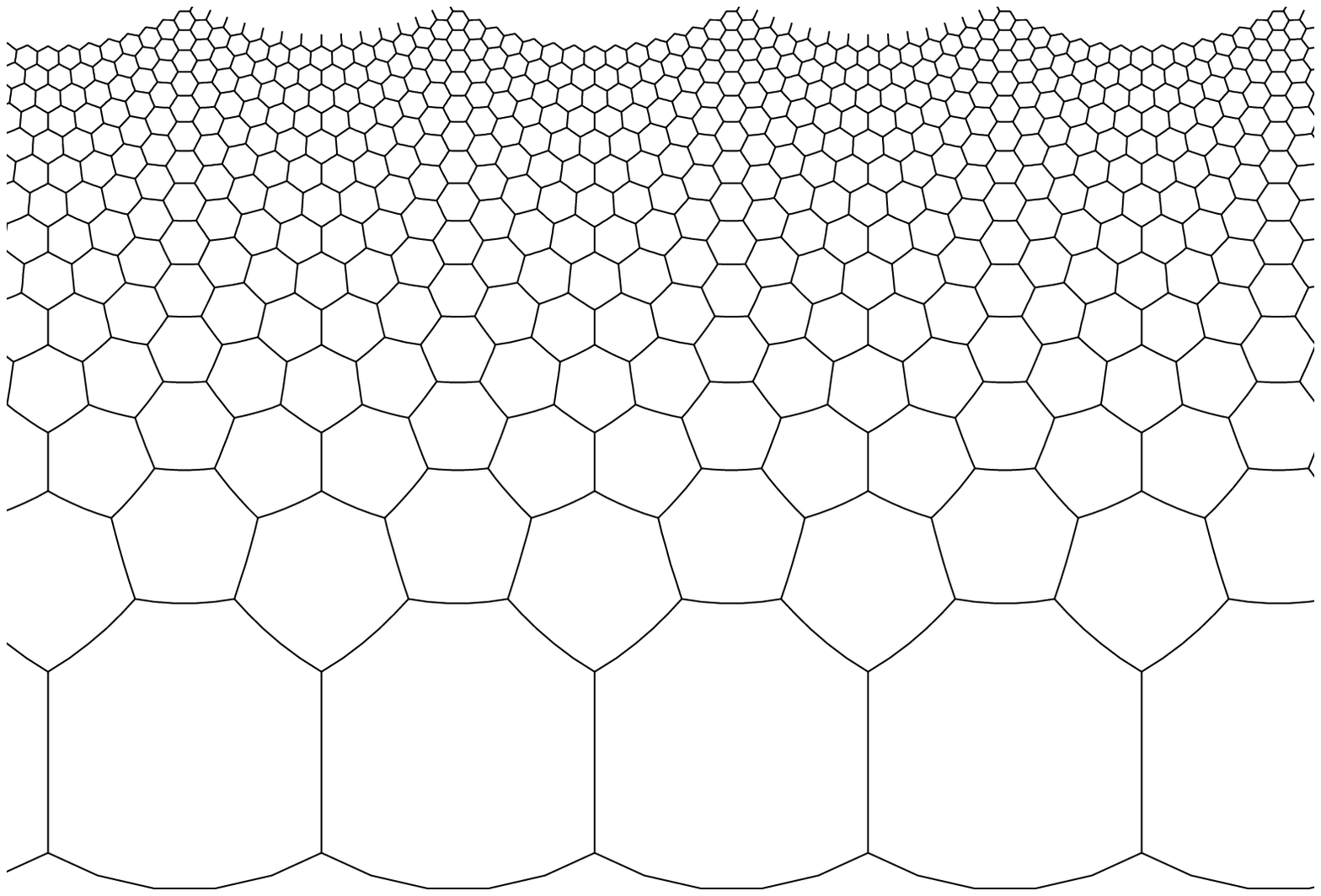}
  \caption{Image of the regular hexagonal foam by the log map.}
  \label{figLog}
\end{figure}

With formula (\ref{eqkn}), we can explicit the edge curvature:
\begin{equation}  \label{eqLogk}
k\simeq |a \tilde \phi| \tilde k + a \cdot n.
\end{equation}
In (\ref{eqLogk}), the first term on the right is the effect of the
local scale change by $|f'|$ ($\tilde k=0$ in a constant pressure
reference, as assumed here); according to the second term, the edges
nearly parallel to $a$ are almost not deformed, up to
similarities, whereas those perpendicular to $a$ are
bent. See fig. \ref{figLog} where $a\propto i$ is vertical.


\subsubsection{Power laws} \label{secMonome}
\begin{figure}  \centering
  \includegraphics[height=5cm]{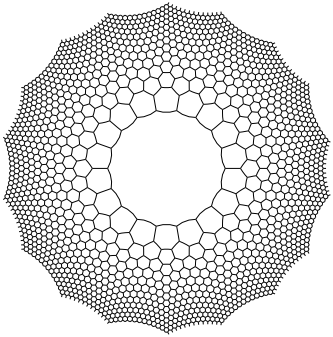}
  \caption{This 18-fold symmetric pattern maps to
    the hexagonal foam of fig. \ref{figHexag} by $f(z)\propto z^3$.}
  \label{figSpherical}
\end{figure}

Upper plates with circular symmetry are described by height functions depending on distance only: $h=h(|z|)$.
For the conformal map, the candidate functions are of the form $f(z)\propto z^\alpha/\alpha$.
An exponent $\alpha = m/6$, $m$ integer, will induce a disinclination of the hexagonal crystal. 
The isobar condition (\ref{eqSolMin}) then implies 
\begin{equation} \label{eqCircular}
h(|z|) \propto |f'(z)| = |z|^{\alpha-1}.
\end{equation}
 For example, a spherical profile corresponds to $\alpha-1=2$, that
 is, to a figure with 18-fold symmetry if the reference foam is the
 regular honeycomb (fig. \ref{figSpherical}). 
This result, derived from constant pressure, differs from 
the prediction of constant volume (\ref{eqConstV}), which is 
$2(\alpha-1)=2$, implying 12-fold symmetry.

\subsection{Discussion, remarks} \label{secRem7}

Equation (\ref{eqfh}) insures compatibility with Laplace law for films in a thin curved interstice. It is interesting to compare it with other conditions.

\subsubsection{Volume conservation versus constant pressure}

In \cite{Drenck}, the set-up was similar to the conventional Hele-Shaw cell, except that the upper plate was not parallel to the bottom one. It was either tilted or moderately curved.
The major constraint, on the mapping, was \emph{volume} conservation in each bubble.
In the long wavelength approximation, equal volume of the reference and the transformed bubbles amounts to the following condition: $h |F'|^2 = \linebreak h |f'|^{-2} = \mathrm{const.}$, or
\begin{equation} \label{eqConstV}
h/h_0 = |f'|^{2}
\end{equation}
for some constant $h_0$.
In \cite{Drenck}, the constant volume predictions do not always match the experiments\footnote{
The clearest evidence is with a spherical upper plate, where constant volume predicts 12-fold symmetry ($\alpha=2$) whereas the sample has 9-fold symmetry.
Possible explanations of this discrepancy may rely in boundary effects,
both near the origin where the liquid occupies a significant region
and at the outer boundary where the absence of the outer part may induce relaxations, and departure from the ideally infinite model.
The experimental protocol followed from the reference to the curved foam may be of importance too.}.

Compared to (\ref{eqConstV}), equation (\ref{eqSolMin}) appears as a constant perimeter condition: at constant film thickness, the amount of liquid per bubble would be the same in the reference and in the transformed froth. This is just an interpretation because (\ref{eqSolMin}) was derived from a constant \emph{pressure} hypothesis and equilibrium.

As already noticed, the predictions deduced from equations (\ref{eqConstV}) and (\ref{eqSolMin}) are different, except for the trivial case where $|f'| \equiv 0$, corresponding to changing the chamber thickness while keeping the two plates parallel, and where (\ref{eqConstV}) and (\ref{eqSolMin}) are compatible.

Let us see the consequences of assuming both constant volume and local film equilibrium.
Inserting (\ref{eqConstV}) into the expression (\ref{eqLapcor}) for the vertical
curvature gives 
\begin{eqnarray} \label{eqkvvconst}
{\scriptstyle \frac{1}{2}} k_v \simeq
 -\mathrm{Re}\left[n\, \partial_z \ln h \right]
 &=& - \mathrm{Re}\left[n\, \partial_z \ln |f'|^2 \right] \nonumber \\
 &=& - \mathrm{Re}\left[n\, \partial_z \ln f' \right]
 = - k
\end{eqnarray}
by (\ref{eqkn}). Therefore the pressure difference (\ref{eqLapcor}) becomes
\begin{equation} \label{eqAntiLaplace}
(P_1-P_2)/ \gamma = k+k_v \simeq -k,
\end{equation}
a result opposite to what a direct readout from the 2D foam and
equations would predict. In figure \ref{figLog}, the pressure \emph{decreases} as one
moves up.

\subsubsection{Short versus long range}
In many cases, a single bubble or liquid bridge between non parallel plates is unstable \cite{liqBridge}. So, for foams, stability is a collective effect; the inner bubbles repel the outer ones.

In the long wave length limit, we can clarify the respective status of the previous equations.
On one hand, with global symmetries such as rotation or unidirectional
translation, bubble volume preservation, equation (\ref{eqConstV}),
essentially forces the positional ordering of the cells with respect
to each other, implying a long range order.
On the other hand, equation (\ref{eqfh}), or its constant pressure version (\ref{eqSolMin}), represents the equilibrium condition for the films; this is a shorter range constraint because the films are small, of the order of the bubble size, and bounded by the Plateau borders. How this effect propagates across film junctions is not clear yet. Neither is it obvious, a priori, that the two conformal maps, in (\ref{eqConstV}) and (\ref{eqfh}), should be the same, despite the fact they were both called $f$.

\subsubsection{Analogy with foams in gravity}

When the film is subjected to a force field $\mu\,\mathbf g$ per unit surface (length in 2D), the force balance in the normal direction is \cite{deGennes}
\begin{equation} \label{eqFbalancePerp}
(P_1-P_2) - 2 \gamma H + \mu\,\mathbf g\cdot\mathbf n = 0.
\end{equation}
The force balance in the tangent direction implies that the surface tension slightly varies with position on the film: $ \gamma= \gamma(\mathbf r)$.
When, moreover, the fluids on both sides of the film are liquids of
mass density $\rho_1, \rho_2$, the local hydrostatic pressure
difference is $(P_1-P_2) = (\rho_1 - \rho_2)\, \mathbf
g\cdot \mathbf r +$ const. .
In particular, if the liquids are the same, as in ferro-fluid foams \cite{Elias}, or if the cells contain gas of negligible weight, then $(P_1-P_2)$ is constant along the film, with a good accuracy.

Comparing (\ref{eqfh}) and (\ref{eqFbalancePerp}) shows that a non parallel profile affects the foam as an effective force field $ \nabla_z(\ln f' - 2\ln h)$.

\section{Conclusion} \label{secConclusion}

When a foam is in contact with a smooth solid surface, the 2D foam formed by the contacts satisfies the 2D Plateau equilibrium conditions in the case of normal incidence.
The case of oblique incidence is more involved and no simple separation appears to occur of the 2D from the 3D problem (Appendix \ref{apNonNormalInc}).

The equilibrium equations for the co-dimension 2 objects (nodes in 2D, Plateau border lines in 3D) are conformal invariant. This has been proved in \cite{ManciniEuf04} and here for foams in Euclidean (flat) spaces.

On the other hand, the equilibrium equation for the co-dimension 1 films or interfaces, namely Laplace-Young's law, is in general not preserved by conformal mappings. For flat 2D foams, the symmetry group of equilibrium patterns is the group of homographies, generated by Euclidean similarities and inversion. In 3D, where the group of conformal transformations reduces a priori to similarities and inversions, Laplace equation is preserved only in the special cases where the cell boundaries are spherical caps.

As an application of the equations at the boundary, we have analysed the films equilibrium conditions for foams enclosed between two non parallel plates. In the limit where the thickness, slope and typical wave vectors of the profile are small, we deduced an approximate map-profile equation which was then solved in the case of constant pressure and illustrated in conditions comparable to available experiments.

In most of the experimental demonstrations of conformality (non parallel plates, spherical cell \`a la Hele-Shaw or ferro-fluids under gravity), the length scale of the mapping is large compared to cell diameter. 
As a small scale condition, our equation completes longer range ones such as the constant volume condition. Moreover, the effect of slightly non parallel plates can be interpreted as an effective force field on the 2D foam.

Examining the projected foam only can be misleading. For example, on
the bottom figure \ref{figLog}, which is the model for a foam
confined in an (approximately) exponential cell as on figure 3(b) of
\cite{Drenck}, the pressure seems to increase as one moves up
(the pressure is higher on the convex side of a curved edge).
But this is wrong.
In these conditions,
solving the equilibrium problem in 3D, taking into account the vertical curvature implied by normal incidence between non parallel plates,
leads to the same image (fig. \ref{figLog}) even at constant pressure;
then, at each edge, the curvature in the vertical direction is exactly opposite to the
horizontal one seen in projection, so that $\Delta P/\gamma = H\simeq 0$.
The same computation at constant volume even predicts that the pressure decreases as one moves up in figure \ref{figLog}.

Our derivation of the map-profile relation relies on an approximation
where the vertical curvature is retained only to the lowest order,
constant. Further investigations are still needed to control this approximation
or to get exact results on foams between non parallel plates.

Experiments at constant pressure, rather than constant volume, would be interesting.
So would be pressure measurements in the experiments at constant volumes.

In the language of defects and Volterra processes, a mapping of the type $z\mapsto z^\alpha$, $\alpha$ rational, represents a disinclination. Now the nearly conformal foams with rotational symmetry also have dislocations, which would be worth taking into account.

\subsection*{Acknowledgement}
\begin{acknowledgement}
We would like to thank D. Weaire for discussions and early
communication of experimental  pictures.
\end{acknowledgement}

\section*{Appendixes}
\begin{appendix}

\section{Planar curves} \label{apCurves}

For a parametrised $C^2$ curve
$ \phi: [t_1,t_2] \rightarrow \mathbb{R}^2$ (or $\mathbb{C}$),\quad $t \mapsto \phi(t)$,
arc length is $\d s = |\d \phi|=|\dot{\phi}| \d t$ and the local
Frenet basis $(\tau, n)$ is composed of $\tau = \frac{\dot{\phi}}{|\dot{\phi}|}$
and $n=i\tau$; $i$ is the imaginary unit.
The scalar curvature is
\begin{equation} \label{eqCurvature}
k =  \frac{n\cdot\ddot{\phi}}{|\dot{\phi}|^2}.
\end{equation}
and the curvature vector
\begin{equation}
k\, n = \frac{\d \tau}{\d s} 
 = \frac{ \ddot{\phi} - (\tau\cdot\ddot{\phi})\; \tau}{|\dot{\phi}|^2}.
\end{equation}

\section{Curvatures} \label{apCurvatures}

We briefly recall the basics on curvatures and curves in surfaces
\cite{Morgan,Dubrovin,Spivak}.

Up to second order around a point $p$ the surface $S$ is the graph of a quadratic form $q(\mathbf a,\mathbf a)$ defined in the tangent plane $T_pS$ with ordinate axis the normal line $\mathbb R\, \mathbf{n}_S(p)$;
$\mathbf n_S$ is the unit normal to $S$.
The corresponding symmetric bi-linear form defines the curvature (linear) operator $Q$ by $q(\mathbf a,\mathbf b)=\mathbf a\cdot Q \mathbf b$ for all $\mathbf a,\mathbf b$ in $T_pS$.

The principal curvatures $\lambda_1, \lambda_2$ and directions $\mathbf u_1, \mathbf u_2$ are the eigenvalues and orthonormal eigenvectors of $Q$. The mean and Gauss curvatures are, respectively, $H=\frac{1}{2} \tr Q = (\lambda_1+ \lambda_2)/2$ and $G= \det Q = \lambda_1 \lambda_2$.

In a local parametrisation chart $(z^1,z^2)\mapsto \mathbf r(z^1,z^2)$, the matrices of the first and second fundamental forms are usually taken as
\begin{eqnarray}
  \mathrm I :& \mathbf r_i dz^i\cdot \mathbf r_j dz^j 
  = g_{ij} dz^i dz^j&\Rightarrow g_{ij} = \mathbf r_i\cdot \mathbf r_j; \label{gij}\\
  \mathrm{II} :& q(\mathbf r_i dz^i, \mathbf r_j dz^j)
  = q_{ij} dz^i dz^j&\Rightarrow q_{ij} = \mathbf r_{ij}\cdot \mathbf n_S,\label{qij}
\end{eqnarray}
where $\mathbf r_i = \partial\mathbf r /\partial z^i$, $i,j=1,2$
and summation is implicitly assumed over repeated indices.
If $Q^i_j$ denotes the matrix defined by $Q\mathbf r_i = \mathbf r_j Q^j_i$, then
$
q_{ij} = \mathbf r_i\cdot Q \mathbf r_j
 = g_{ik} Q^k_j
$
or, as $2\times2$ matrices,
\begin{equation} \label{q=gQ}
  Q = g^{-1} q.
\end{equation}
The identity $\mathbf n_S \cdot\mathbf a=0$ for all $\mathbf a$ in $T_pS$ implies 
$\mathbf n_{S,i} = -\mathbf r_j Q^j_i$.

The sectional curvature of $S$ in the direction $\mathbf u \in T_pS$ is
\begin{equation} \label{defSectional}
  k_n(\mathbf{u}, S)=\mathrm{II}(\mathbf{u},\mathbf{u})=q(\mathbf{u},\mathbf{u})
\end{equation}
Completing $\mathbf u$ into an orthonormal basis $(\mathbf u, \mathbf v)$ and calling $\alpha$ the angle between $\mathbf u_1$ and $\mathbf u$, the change of coordinates implies
(Euler's Theorem)
\begin{equation}\label{eqSectional}
  k_n(\mathbf{u}, S)=\mathbf{u}\cdot Q\mathbf{u}
  = \lambda_1 \cos^2\alpha+\lambda_2\sin^2\alpha,
\end{equation}
or, since $\lambda_1, \lambda_2$ are the solutions of the characteristic equation
$\lambda^2 -2H\lambda + G = 0$,
\begin{equation} \label{eqSectionHG}
  k_n(\mathbf{u}, S) = H + (H^2-G)^{1/2} \cos(2\alpha).
\end{equation}
Finally, recall the invariance of the trace: 
\begin{equation} \label{eqTrace}
  2H
  =\lambda_1 +\lambda_2 = \mathbf{u}\cdot Q\mathbf{u}+\mathbf{v}\cdot Q\mathbf{v}
  = k_n(\mathbf{u}, S)+k_n(\mathbf{v}, S)
\end{equation}
for any orthonormal pair $(\mathbf u, \mathbf v)$ of tangent vectors.

When a curve $\phi$ (parametrised by arc length $s$) is embedded in a surface $S$, its tangent vector $\dot{\mathbf\phi}(s)$ is, of course, in the tangent plane $T_{\phi(s)}S$. Its curvature vector $\mathbf k = \ddot{\mathbf\phi}(s)$, always perpendicular to the curve, may be decomposed into 
\begin{equation} \label{eqcurvcurv}
  \mathbf{k}= \mathbf{k}_g + \mathbf{k}_n = 
  k_g(\phi, S)\, \mathbf{n}_S \wedge \dot\mathbf\phi + k_n(\phi, S)\, \mathbf{n}_S
\end{equation}
where 
$\mathbf{b}_S =\mathbf{n}_S \wedge \dot\mathbf\phi$ is the co-normal. The component $k_g(\phi, S)$ tangent to the surface is the geodesic curvature of $\phi$ in $S$; the other is the normal curvature $k_n(\phi, S)$ which coincides with the sectional curvature of $S$ in the direction $\dot\mathbf\phi$: $k_n(\dot\mathbf\phi, S)$.

\section{Bubble-glass contacts with different surface tensions}\label{apNonNormalInc}

When the surface tension along the solid surface differs from cell to cell, the incidence is not normal anymore. Let us treat this case.

\subsection{Equilibrium equation for non normal incidence} \label{secGenerinc}

Consider, again, an equilibrated triple $(\psi, \mathcal{F}_1, \mathcal{F}_2, \mathcal{F}_3)$
Each of the contact angles $\theta_j = \angle(\mathbf n_j, \mathbf n_S), j=1,2,3$, satisfies Young's law (\ref{young}), the force balance projected tangentially to the surface.
The component along the normal axis $\mathbf n_S$ involves the normal reaction of the glass, which is an additional unknown.

\begin{figure} \centering
  \includegraphics[width=7.5cm]{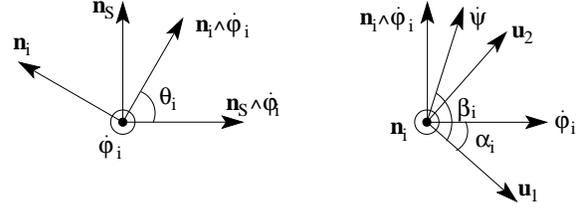}
  \caption{Relevant vectors in $\dot\phi_{j\perp}$ (left) and in $T_p\mathcal F_j$ (right).}
  \label{figBasis}
\end{figure}

If $\phi_j$ is the contact curve $\mathcal{F}_j \cap \mathcal{S}$, $\dot{\phi}_j$ is normal to both $\mathbf n_j$ and $\mathbf n_S$ (fig. \ref{figBasis}), so that
\begin{equation} \label{prodvettnuo}
  \mathbf n_S \cdot\mathbf n_j = \cos\theta_j,  \qquad
  \mathbf n_S\wedge\mathbf n_j=\dot{\phi}_j\sin(\theta_j).
\end{equation}  
The following equations follow from taking the dot and cross products of $\mathbf n_S$ with Plateau's law (\ref{platoPerp}):
\begin{eqnarray}
  \label{somcos} &\sum_{j=1}^3\gamma_j\cos(\theta_j)=0,\\
  \label{plateaunuovo} &\sum_{j=1}^3\gamma_j\dot{\phi}_j\sin(\theta_j)=0.
\end{eqnarray}

Regarding the curvatures, we shall prove the following: at the triple point $p=\psi\cap\mathcal S$,
\begin{eqnarray} \label{eqfinal}
\sum_j\gamma_j\left[ k_n(\phi_j, \mathcal S)\cos\theta_j
  - k_g(\phi_j, \mathcal S)\sin\theta_j\right] \nonumber \\
  = \sum_j\gamma_j (H_j^2-G_j)^{1/2} \cos(2\alpha_j).
\end{eqnarray}
It involves the angle $\alpha_j=\angle(\mathbf u_1(\mathcal{F}_j), \dot\phi_j)$ between the curve $\phi_j$ and the first principal direction of $\mathcal{F}_j$.

Equations (\ref{plateaunuovo}), (\ref{eqfinal}) are the analogous of (\ref{2Dplatot}), (\ref{2Dplatok}). In the present case where the films are not perpendicular to the solid surface, the equations in the boundary do not decouple from the full problem of the 3D foam.

\subsection{Derivation of the equilibrium equation (\ref{eqfinal})} \label{pfGenerinc}

Let $(\mathbf u_1, \mathbf u_2)=(\mathbf u_1(\mathcal{F}_j),\mathbf u_2(\mathcal{F}_j))$ be the principal directions of $\mathcal F_j$. We use the following angles:
$\alpha_j=\angle(\mathbf u_1, \dot\phi_j),\ \beta_j=\angle(\mathbf u_1, \dot\psi),\ \theta_j=\angle(\mathbf n_S, \mathbf n_j)$;
\begin{equation} \label{phiUmat}
  (\begin{array}{lr} \dot\phi_j&\dot\psi\end{array}) =
  (\begin{array}{lr}\mathbf u_1&\mathbf u_2\end{array})
  \left(\begin{array}{lr} \cos\alpha_j&\cos\beta_j\\
      \sin\alpha_j&\sin\beta_j\end{array}\right).
\end{equation} 

As both $\phi_j$ and $\psi$ are curves in $\mathcal F_j$, equation (\ref{eqSectionHG}) reads
\begin{eqnarray}
  \label{eqKnphi} &k_n(\dot\phi_j, \mathcal F_j) = H_j + (H_j^2-G_j)^{1/2} \cos(2\alpha_j),\\
  \label{eqKnpsi} &k_n(\dot\psi, \mathcal F_j) = H_j + (H_j^2-G_j)^{1/2} \cos(2\beta_j).
\end{eqnarray}
Putting the coefficients $\gamma_j$ and summing (\ref{eqKnphi}, \ref{eqKnpsi}) over the triple around $\psi$ gives
\begin{eqnarray}
  \label{eqsumKnphi} \sum_j\gamma_j k_n(\dot\phi_j, \mathcal F_j) = \sum_j\gamma_j (H_j^2-G_j)^{1/2} \cos(2\alpha_j),\qquad\\
  \label{eqsumKnpsi} \sum_j\gamma_j k_n(\dot\psi, \mathcal F_j) = \sum_j\gamma_j (H_j^2-G_j)^{1/2} \cos(2\beta_j)=0.\quad
\end{eqnarray}
The second vanishes by (\ref{platoKn}).
In the first, (\ref{eqlemma}) allows to express the LHS in terms of curvatures in $\mathcal S$:
\begin{eqnarray} \label{eqfinal2}
  {\textstyle \sum_j}\gamma_j\left\lbrack k_n(\phi_j, \mathcal S)\cos\theta_j -
    k_g(\phi_j, \mathcal S)\sin\theta_j \right\rbrack \nonumber \\
  = {\textstyle \sum_j}\gamma_j (H_j^2-G_j)^{1/2} \cos(2\alpha_j).
\end{eqnarray}
This is (\ref{eqfinal}).
\qed

\subsection{Remark}

The equations for the case of normal incidence follow from the general ones. Indeed, $|\theta_j|=\pi/2 = |\beta_j-\alpha_j|$ implies $\cos(2\beta_j)=-\cos(2\alpha_j)$; then, in (\ref{eqfinal}), the $\cos\theta_j$ term disappears from the LHS and the RHS is zero by (\ref{eqsumKnpsi}), so that 
(\ref{eqfinal}) reduces to (\ref{2Dplatok}).

\end{appendix}

%

\begin{thebibliography}{9}
%
%

\bibitem{WeaireRiv}
Weaire D. and Rivier N.,Contemp. Phys. \textbf{25}, (1984) 59

\bibitem{WeaireHut}
Weaire~D. and Hutzler~S., \textit{The physics of foams} 
(Clarendon Press, Oxford, 1999)

\bibitem{WeaireInstab}
Cox S.J., Weaire D., Vaz M.F., Eur. phys. J. E \textbf{7}, (2002) 311-315

\bibitem{Brakke}
Brakke K., Morgan F., Eur. phys. J. E \textbf{9}, (2002) 453-460

\bibitem{Fomenko}
A.T. Fomenko, \textit{ The Plateau problem} 
(Gordon and Breach, 1990)

\bibitem{DiMeglioSenden}
DiMeglio J.M., Senden T., Communication at \textit{Eufoam04}, (2004), Marne la Vallee, F, July 2004

\bibitem{Drenck}
W.~Drenckhan, D.~Weaire and S.J.~Cox, Eur. J. Phys. \textbf{25}/3, (2004) 429-438

\bibitem{Moukarzel}
Moukarzel C., Phys. Rev. E \textbf{55}, (1997) 6866--6880

\bibitem{Weaire}
Weaire D., Phil. Mag. Lett. \textbf{79}, (1999) 491--495

\bibitem{ManciniO}
Mancini M. and Oguey C., Phil. Mag. Lett.  \textbf{83}, (2003) 643-649

\bibitem{Morgan}
Morgan F., \textit{Riemannian geometry : a beginner's guide} 
(Wellesley, MA : A.K. Peters, 1998)

\bibitem{Willmore}
T.\,J.~Willmore, \textit{Riemannian geometry} 
(Clarendon Press, Oxford, 1996)

\bibitem{Dubrovin}
B.A. Dubrovin, A.T. Fomenko, S.P. Novikov, \textit{Modern geometry --
  methods and applications. Part 1. The geometry of surfaces,
  transformation groups, and fields.} 
(Springer-Verlag, Berlin, 1992), \S\S~5, 8 \&\ 15

\bibitem{ManciniEuf04}
Mancini M. and Oguey C., submitted to Col Surf A (2004)

\bibitem{RothenPierRivJo} 
F.~Rothen, P.~Pieranski, N.~Rivier and A.~Joyet, Eur.~J.~Phys. \textbf{14}, (1993) 227

\bibitem{RothenPier}
Rothen F, Pieransky P, Phys. Rev. E \textbf{53}, (1996) 2828-2842

\bibitem{Elias}
Elias~F., Bacri~J.C., de Mougins~H., Spengler~T., Phil. Mag. Lett. \textbf{79}, (1999) 389-397

\bibitem{RivierConf}
N.~Rivier, D.~Reinelt, F.~Elias and C.~Vanden Driessche,
in
\textit{Proceedings of the International Workshop on Foams and Films,
  Leuven (Belgium), 5-6 March 1999}. Editors D.~Weaire and J.~Banhart 
(Verlag MIT publish, 1999)

\bibitem{liqBridge}
Paul Concus and Robert Finn, Phys. Fluids, \textbf{10}(1), (1998) 39-43

\bibitem{deGennes}
P.G. de Gennes, F. Brochard, D. Que\'re\', \textit{Gouttes, bulles,
  perles et ondes} 
(Belin, Paris, 2002)

\bibitem{Spivak}
M. Spivak, \textit{A Comprehensive Introduction to Differential
  Geometry} 
(Publish or perish, Houston, 1979)


\end{thebibliography}
%

\end{document}